\documentclass[twocolumn,showkeys,showpacs,preprintnumbers,amsmath,amssymb,prl,superscriptaddress]{revtex4}
\usepackage{amsmath}
\usepackage{graphicx}
\usepackage{dcolumn}
\usepackage{bm} 

\begin{document}
\title{\boldmath Spontaneous formation of a superconducting and
antiferromagnetic hybrid state in SrFe$_2$As$_2$ under high pressure}
\author{K.~Kitagawa}
\email{kitag@issp.u-tokyo.ac.jp}
\affiliation{Institute for Solid State Physics, University of Tokyo, Kashiwanoha, Kashiwa, Chiba 277-8581, Japan}
\author{N.~Katayama}
\altaffiliation{Present address: Department of Physics, University of Virginia,
Charlottesville, Virginia 22904, USA}
\affiliation{Institute for Solid State Physics, University of Tokyo, Kashiwanoha, Kashiwa, Chiba 277-8581, Japan}
\author{H.~Gotou}
\affiliation{Institute for Solid State Physics, University of Tokyo, Kashiwanoha, Kashiwa, Chiba 277-8581, Japan}
\author{T.~Yagi}
\affiliation{Institute for Solid State Physics, University of Tokyo, Kashiwanoha, Kashiwa, Chiba 277-8581, Japan}
\author{K.~Ohgushi}
\affiliation{Institute for Solid State Physics, University of Tokyo, Kashiwanoha, Kashiwa, Chiba 277-8581, Japan}
\affiliation{JST, TRIP, 5 Sanbancho, Chiyoda, Tokyo 102-0075, Japan}
\author{T.~Matsumoto}
\affiliation{Institute for Solid State Physics, University of Tokyo, Kashiwanoha, Kashiwa, Chiba 277-8581, Japan}
\author{Y.~Uwatoko}
\affiliation{Institute for Solid State Physics, University of Tokyo, Kashiwanoha, Kashiwa, Chiba 277-8581, Japan}
\affiliation{JST, TRIP, 5 Sanbancho, Chiyoda, Tokyo 102-0075, Japan}
\author{M.~Takigawa}
\affiliation{Institute for Solid State Physics, University of Tokyo, Kashiwanoha, Kashiwa, Chiba 277-8581, Japan}
\affiliation{JST, TRIP, 5 Sanbancho, Chiyoda, Tokyo 102-0075, Japan}

\date{\today}
\begin{abstract}
We report a novel superconducting (SC) and antiferromagnetic (AF) hybrid state
in SrFe$_2$As$_2$ revealed by {$^{75}$As} nuclear magnetic
resonance (NMR) experiments on a single crystal under highly hydrostatic
pressure up to 7~GPa. The NMR spectra at 5.4~GPa indicate simultaneous development of the SC and AF orders below 30~K.
The nuclear spin-lattice relaxation rate in the SC domains shows a substantial
residual density of states, suggesting proximity effects due to spontaneous
formation of a nano-scale SC/AF hybrid structure. This entangled behavior is a
remarkable example of a self-organized heterogeneous structure in a clean system.
\end{abstract}

\maketitle
Competition between magnetism and superconductivity is ubiquitous in
unconventional superconductors such as cuprates and heavy
fermions\cite{UemuraNM}, as is the case for iron
pnictides\cite{UemuraNM,IshidaJPSJReview}.
Precise determination of the phase diagram, in particular whether the
antiferromagnetic (AF) and the superconducting (SC) phases mutually exclude or
coexist, is often hampered by compositional and structural inhomogeneity, which are
inevitable when the SC state is induced by carrier
doping\cite{ParkKdopePRL,PrattCodopePRL}.
 
While this is the case for cuprates, there is another route to the SC state in
iron pnictides.  SrFe$_2$As$_2$ shows a phase transition
from the tetragonal paramagnetic (PM) state into the orthorhombic AF state below 200~K at
ambient pressure\cite{KanekoSr122NS,KitagawaSr122NMR}. It is reported to become SC
not only by chemical
substitution\cite{SasmalSrK122,JasperSrCo122}
 to change carrier concentration but also by applying sufficiently high pressure to suppress the AF
state\cite{Alireza122HP,KotegawaSr122HP,Matsubayashi122HP}.
The pressure-induced SC state provides an opportunity to unravel intrinsic 
behavior caused by interplay between the AF and SC order unaffected by
disorder. This
motivated us to perform high-pressure NMR spectroscopy, a powerful tool
to examine coexistence of the SC and AF states.

We developed a new type of opposed-anvil pressure cell\cite{KitagawaHPCell},
which enables NMR experiments in quasi-hydrostatic pressure up to 9.4~GPa by using argon as the pressure
transmitting medium\cite{TateiwaMedia}. The cell was mounted on a geared
double-axis goniometer allowing precise sample alignment within one degree.
The single crystals of SrFe$_2$As$_2$ were prepared by
self-flux method. 
A small piece ($1\times 0.7\times 0.15$~mm$^3$) was cut from the sample used in the
study at ambient pressure\cite{KitagawaSr122NMR}. NMR signals from
Sn and Pt foil were used to determine pressures in
situ\cite{KitagawaHPCell}.

\begin{figure}[htbp]
\centering
\includegraphics[width=0.9\linewidth]{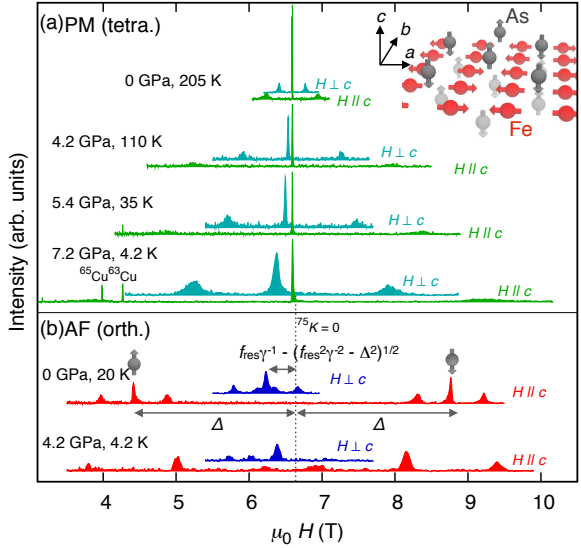}
\caption{(Color online) {$^{75}$As}-NMR spectra of 48.31~MHz at
various $P$. (a) The spectra in the PM tetragonal normal state
(slightly above $T_\text{N}$ or $T_\text{c}$) for $H \parallel c$
(green) and $H \perp c$ (sky blue). 
Although the central line for $H \perp c$ becomes
broader at higher $P$ due to distribution of the second-order quadrupole shift,
it remains sharp up to 7.2~GPa for $H \parallel c$. The vertical dashed line represents the
unshifted resonance position $f_\text{res}\gamma^{-1}$ ($= 6.63$~T). (b) The
spectra in the AF state for $H \parallel c$ (red) and $H \perp c$
(blue). The gray arrows indicate the splitting ($H \parallel c$) or the shift
($H \perp c$) induced by the hyperfine fields $\pm \Delta$ along the $c$-axis,
transferred from the stripe-type AF Fe moments [illustrated at the right-top corner in (a)]. There are four satellite
lines for $H \perp c$, indicating the twinned orthorhombic structure with asymmetric electric
field gradients. }
\label{fig:spectra-pmaf}
\end{figure}
\begin{figure}[htbp]
\centering
\includegraphics[width=0.9\linewidth]{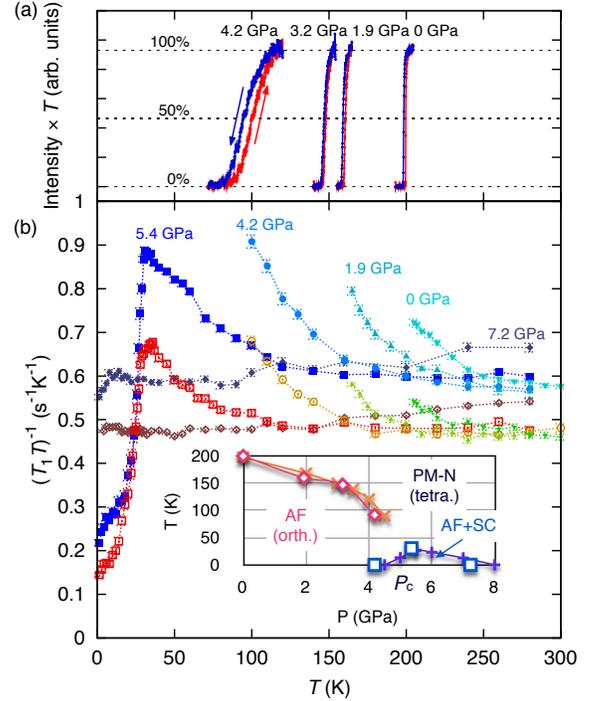}
\caption{(Color online) (a) The AF transition
detected by tracking the peak intensity of the PM central line for
$H \parallel c$.
The rates for the $T$ sweep were kept slower than 0.05, 0.1, 0.3~K/min. for
1.9, 3.2, 4.2~GPa, respectively. Sharp transition with hysteresis demonstrates good
homogeneity of the sample and the first-order AF transition. (b) $(T_1T)^{-1}$
in the PM state for $H \perp c$ (the plots in dark blue to sky blue with higher values)
and for $H \parallel c$ (the lower plots in brown
to green). The inset shows the $P$-$T$ phase diagram. $T_\text{N}$ is determined
from the 50\% recovery of the PM intensity on warming in (a) (open diamonds).
The bulk SC transition (open squares) is detected in
our NMR experiment only at 5.4~GPa. $T_\text{c}$
is determined from the peak in $(T_1T)^{-1}$ at 11~T for $H \perp c$. The SC
transition is absent below 4.2~GPa and at 7.2~GPa. Crosses represent the data
reported based on the resistivity (for AF) and the ac susceptibility (for SC)
measurements\cite{Matsubayashi122HP}. 
}
\label{fig:t1}
\end{figure}
Figure~\ref{fig:spectra-pmaf} shows the {$^{75}$As}-NMR
spectra in the PM and AF states at various $P$. The
quadrupole interaction splits the resonance of {$^{75}$As} nuclei with spin $I = 3/2$ into three lines
corresponding to the transitions $I_z = m \leftrightarrow m - 1$ $(m = \pm 1/2,3/2)$ at the frequencies 
$f_\text{res} = \gamma|\bm{H} + \bm{H}_\text{hf}| + (m - 1/2)\nu +
\text{(second-order quadrupole shift)}$, where $\gamma = 7.2902$~MHz/T is the
gyromagnetic ratio, $H_\text{hf}$ is the hyperfine field from the Fe spins, and $\nu$ is
proportional to the electric field gradient along the field direction. In the
PM state, $H_\text{hf}$ is uniquely given and proportional to the external
field, $H^\alpha_\text{hf} = K^\alpha H^\alpha$, where $K^\alpha$ is the Knight
shift along the $\alpha$ direction. In the AF state the spectra for $H
\parallel c$ split symmetrically into two sets of three lines. However, the
central line $(m = 1/2)$ for $H \perp c$ does not split, indicating that the
hyperfine fields are parallel to the $c$-axis, $H_\text{hf} = (0, 0,
\pm\Delta)$. This is compatible with the commensurate stripe AF structure with
$\bm{q} = [101]$ and the AF moments parallel to the $a$-axis as previously
reported\cite{KitagawaSr122NMR,KitagawaBa122}. Our results prove that the
ground state below 4.2~GPa has the stripe-type AF order with the orthorhombic
structure.

The AF transition temperature $T_\text{N}$ can be determined from
the vanishing of the PM spectral intensity [Fig.~\ref{fig:t1}(a)]. The
hysteresis provides clear evidence for a first-order transition. Below
4.2~GPa, no fraction of the PM state remains at low $T$. Figure~\ref{fig:t1}(b)
shows $T$-dependence of the nuclear spin-lattice relaxation rate
$T_1^{-1}$ divided by $T$. $(T_1T)^{-1}$ shows an upturn as $T$
approaches $T_\text{N}$ below 4.2~GPa, indicating development of AF spin
fluctuations in the PM state even though the transition is
first-order. At 5.4~GPa, an increase of $(T_1T)^{-1}$ is also observed above
30~K, followed by a sudden decrease at lower $T$. As we discuss later, the PM spectrum does not vanish at low $T$ at
5.4~GPa and there is sufficient evidence for a bulk SC state. The reduction
of $(T_1T)^{-1}$ is then a consequence of opening of the energy gap in the SC
state. When $P$ is increased to 7.2~GPa, $(T_1T)^{-1}$ is nearly
independent of $T$, exhibiting neither an SC transition nor development of AF
fluctuations. 
\begin{figure*}[htbp]
\centering
\begin{minipage}{0.48\textwidth}
\centering
\includegraphics[width=0.95\linewidth]{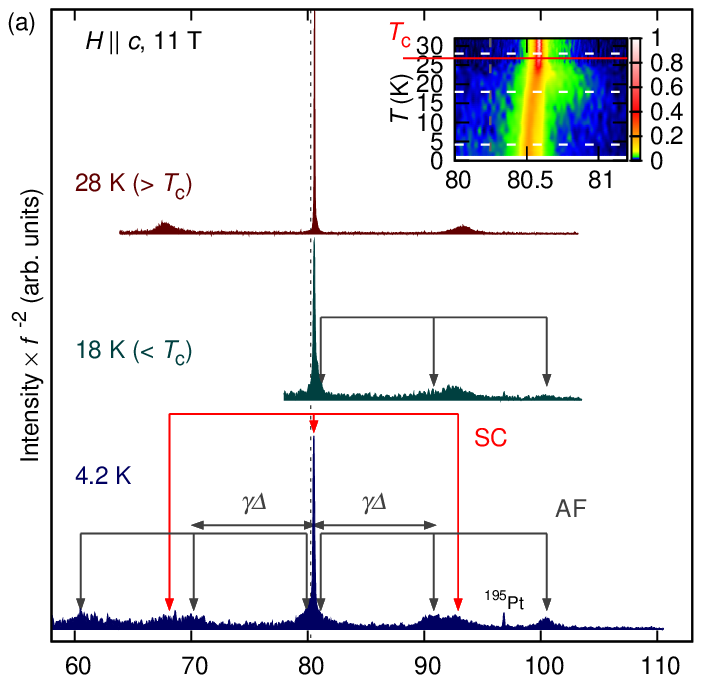}
\includegraphics[width=0.95\linewidth]{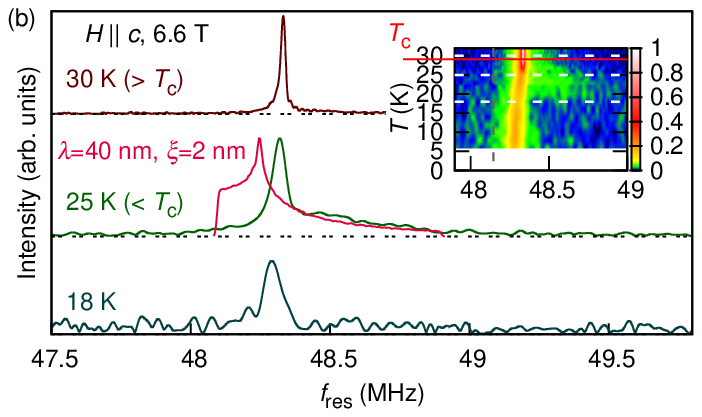}
\end{minipage}
\begin{minipage}{0.48\textwidth}
\centering
\includegraphics[width=0.95\linewidth]{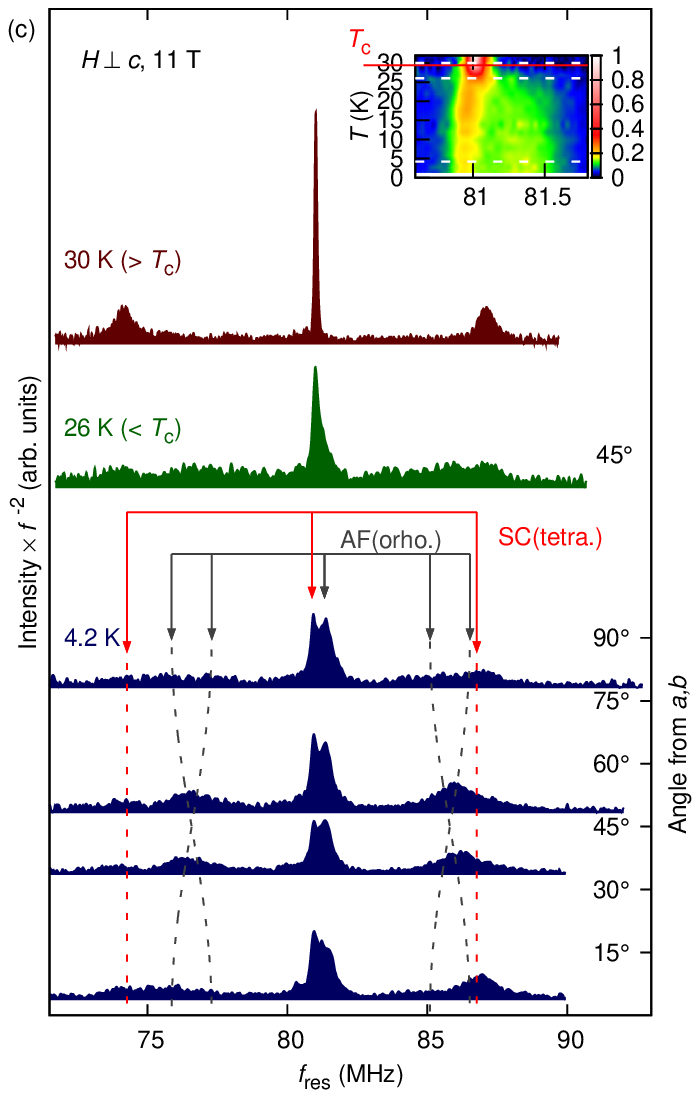}
\end{minipage}
\caption{(Color) {$^{75}$As}-NMR spectra at 5.4~GPa showing the
coexistence of the AF and SC states. The insets show the color plots indicating
the $T$-dependence of the spectral intensity distribution near the central line.
The
white dashed lines indicate the $T$ values of the spectra in the main panels. 
(a) The NMR spectra for $H \parallel c$ at 11~T
measured at 28~K (above $T_\text{c}$), 18~K (below $T_\text{c}$), and 4.2~K.
At 4.2~K, the spectrum shows contributions both from the
commensurate stripe-type AF domains (gray arrows) and from the PM-SC domains (red arrows).
(b) The PM (SC or normal) central lines for $H \parallel c$ at 6.6~T.
Anomalous line shape with asymmetric broad tail is observed for $18~\text{K} < T <
T_\text{c}$ (see the inset). Standard models of vortex lattices are unable to fit the spectrum at
25~K even if an extremely short penetration depth $\lambda$ is assumed. An
example is shown by the red line calculated with the modified London model for a
rectangular vortex lattice\cite{YaouancVortex,NakaiVortex,CurroHFReview}.
(c) The spectra for $H \perp c$ at 11~T. At
4.2~K, the satellite lines from the AF domains split into two sets. The
azimuth angle dependence of the satellite positions in the $ab$-plane are
reproduced by the asymmetry parameter $|(\nu_a - \nu_b)/\nu_c|= 0.15$ as
shown by the dashed lines. This proves the twinned orthorhombic structure of
the AF domains. The broadened spectrum at 26~K from the AF domains indicates
modulation of AF moments and the structure, most likely an incommensurate
spin-density-wave type.}
\label{fig:spectra-sc}
\end{figure*}
\begin{figure}[htbp]
\centering
\includegraphics[width=0.9\linewidth]{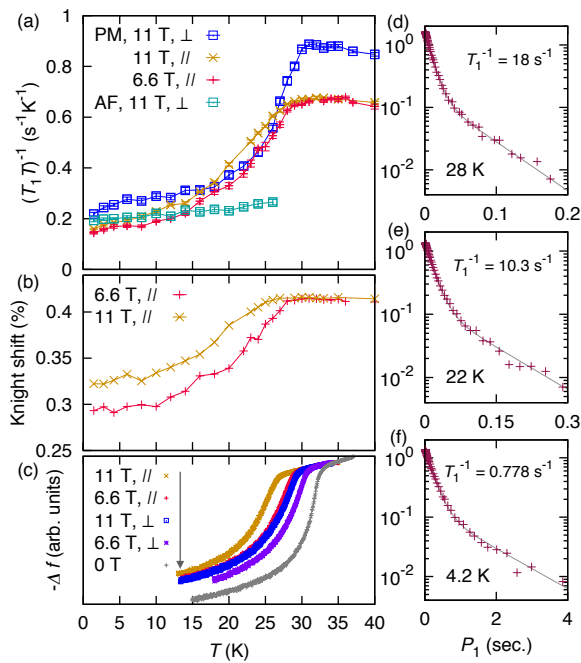}
\caption{(Color online) Evidence for bulk SC state at
5.4~GPa. (a) The $T$-dependence of $(T_1T)^{-1}$ in the PM state
for $H \perp c$ at 11~T (blue) and for $H \parallel c$ at 11~T (orange) and at 6.6~T (red).  The blue
and the red data are identical to those shown in Fig.~\ref{fig:t1}(b). Opening
of the SC gap causes sudden decrease of $(T_1T)^{-1}$ below $T_\text{c}$.
Also shown are $(T_1T)^{-1}$ data in the
AF state for $H \perp c$ at 11~T (sky blue). All
measurements were made on the central line, whose positions are shown by the
arrows in Fig.~\ref{fig:spectra-sc}.
(b) The $T$-dependence of the Knight shift for $H \parallel c$ at 6.6~T and
11~T. The reduction below $T_\text{c}$ indicates the spin-singlet pairing in the SC state.
(c) The ac susceptibility
measured  by the self resonance method using the NMR coil. The change of the
resonating frequency $-\Delta f$ represents the Meissner shielding.
(d)--(f) Typical recovery curves of the nuclear magnetization during the $T_1$
measurement at 28~K (immediately above $T_\text{c}$), 22~K, and 4.2~K (below
$T_\text{c}$) for $H \parallel c$ at 11~T.  Each curve is fit very well to the
formula for spin $3/2$ nuclei, $\{ \exp(-t/T_1) + 9 \exp(-6t/T_1)\}/10$, with a
single value of $T_1$, supporting uniform relaxation process.}
\label{fig:t1-sc}
\end{figure}

Let us examine the microscopic features of the SC state at
5.4~GPa. Figure~\ref{fig:spectra-sc} shows the NMR spectra
obtained by sweeping frequency at fixed fields\footnote{
Each Fourier spectrum of an echo is summed while the frequency is being swept
[Fourier-step-summing (FSS)]. To take
spectrum over a very wide frequency range, we superpose the FSS power spectra 
by shifting the center
frequency by 1~MHz and readjusting the electric circuit and transmitting power.}.
The spectra immediately above the SC transition temperature $T_\text{c}$ show identical features to those at
35~K [Fig.~\ref{fig:spectra-pmaf}(a)], and hence belong to the PM and normal (N) state. The central line for $H
\parallel c$ broadens suddenly below $T_\text{c}$
[see the insets and Fig.~\ref{fig:spectra-sc}(b)] due to field distribution in
the SC mixed state. At 4.2~K, the spectra clearly show coexistence of the
tetragonal PM-SC and the orthorhombic AF domains with comparable volume fractions as indicated in
Figs.~\ref{fig:spectra-sc}(a) and (c). The broad
asymmetric central line for $H \perp c$ at 26~K [Fig.~\ref{fig:spectra-sc}(c)]
indicates that the AF domains appear immediately below $T_\text{c}$. Thus only
one phase transition from the N-PM state to the SC/AF coexisting state occurs
over the entire sample. This would not be possible if the coexistence were due
to inhomogeneity. Instead, the results strongly suggest formation of a
self-organized SC/AF hybrid structure.

The $T$-dependence of $(T_1T)^{-1}$ and
the Knight shifts were measured at 5.4~GPa [Figs.~\ref{fig:t1-sc}(a) and (b)].
Both exhibit sudden decrease due to opening of the SC gap with spin-singlet
pairing below $T_\text{c} = 27\text{--}30$~K, depending slightly on the field
values. The values of $T_\text{c}$ agree well with the onset of the diamagnetic
response [Fig.~\ref{fig:t1-sc}(c)].
These results provide definitive evidence for the bulk SC state. What is anomalous though is that
$(T_1T)^{-1}$ approaches a finite value towards 0~K, indicating a finite
density of state (DOS), i.e. gapless superconductivity. Moreover, the SC and
AF domains show nearly the same value of $(T_1T)^{-1}$ at low $T$. To
demonstrate good electronic homogeneity, the recovery curves in the relaxation
measurements are shown in Figs.~\ref{fig:t1-sc}(d)--(f) at several $T$. All curves are fit
extremely well by assuming a single value of $T_1^{-1}$.
Since even minor
disorder often causes significant distribution in $T_1^{-1}$, this is a
stringent proof for good homogeneity. Therefore, the coexistence of SC/AF
domains and the residual DOS in the SC domains are intrinsic and free from
material inhomogeneity.
The origin of the residual DOS should be one of the following: (i) pair breaking
due to disorder or magnetic field, (ii) existence of gapless Fermi surfaces
not involved in the SC pairing, or (iii) proximity effect from nearby AF
domains. We can exclude the case (i) based on the nearly field-independent and
uniform behaviour of $(T_1T)^{-1}$. Since the gapless behaviour persists down
to 1.5~K, orders of magnitude smaller than $T_\text{c}$, the case (ii) alone
is also unlikely. 
Nevertheless,
proximity effects from the AF domains (case iii) can squeeze small gaps on parts of the Fermi surfaces.
Note that the size of SC domains must be sufficiently small in order for
proximity to be effective. This suggests spontaneous formation of nano-scale
periodic structure of alternating SC and AF domains.

We now add the SC phase boundary
to the $P$-$T$ phase diagram in the inset of Fig.~\ref{fig:t1}(b). The SC/AF
hybrid state appears only in a very narrow $P$-region over 5~GPa. It is reported that
improved hydrostaticity results in higher critical pressure 
$P_\text{c}$ for the appearance of the SC state\cite{KotegawaUniaxial}. Our
value $P_c \sim 5$~GPa is higher than the previous results, indicating good
hydrostaticity of argon medium. The narrow $P$-range of the SC state is in
marked contrast to the stable pure SC state obtained by chemical
doping\cite{KamiharaJACS,RotterBK122,SasmalSrK122,JasperSrCo122}, and points to
a rather fragile nature of the SC/AF hybrid state. Note that the equal numbers
of electrons and holes are preserved under pressure but not by doping.

Next, we focus on the $T$-dependence of the spectra at 5.4~GPa immediately below
$T_\text{c}$. We first discuss the spectra from the AF domains. For $H
\parallel c$, the spectral intensity from the commensurate AF domains is
reduced with increasing $T$ and only weakly visible at 18~K
[Fig.~\ref{fig:spectra-sc}(a)]. On the other hand, for $H \perp c$, the AF
spectra do not lose intensity but broaden near $T_\text{c}$. The asymmetric
broadening of the central line and the much wider satellites of the spectrum
at 26~K [Fig.~\ref{fig:spectra-sc}(c)] 
indicate incommensurate modulation in both the AF moments and the structure.
Since even the second order effect of $H_\text{hf}$ for $H \perp c$ causes 
substantial broadening, the first order effects should easily wipe out the AF
spectra for $H \parallel c$, explaining the loss of intensity.
Our results then suggest a crossover at $T^* \sim 18$~K from
the low-$T$ stripe AF state with a uniform magnitude of moments to the
high-$T$ modulated (incommensurate) structure.

In the same $T$ region $(T^* < T < T_\text{c})$, we observed
anomalous spectral shape as shown in
Fig.~\ref{fig:spectra-sc}(b). The central line at 25~K for $H \parallel c$
consists of a moderately broadened line on top of a much broader tail. While
the former persists to low $T$, the latter disappears below $T^*$.
Since similar values of $T_1^{-1}$ are obtained on and off the main peak,
they both belong to the SC-PM domain. Note that $T_1^{-1}$ for the
AF domain is smaller by a factor of three at 25~K [Fig.~\ref{fig:t1-sc}(a)].
The width of the broad tail decreases slightly with increasing field, indicating that the
origin is related to the SC diamagnetic current. Nevertheless, the entire line
shape cannot be explained by standard vortex lattices with any choice of
parameters (see the red line). The puzzling line shape above $T^*$ indicates
anomalous modulation of the SC diamagnetic current with a much shorter length
scale than the penetration depth, where modulation of the AF moments also
appears.

Coexistence of AF and SC states has been observed in some heavy
fermion compounds. For example, CeRhIn$_5$ under pressure shows microscopic
coexistence of the AF and SC orders, which are spatially
indistinguishable\cite{CurroHFReview,KawasakiCeRhIn5}. In
CeIn$_3$ the AF and SC orders occur in different spatial region at different
transition temperatures\cite{CurroHFReview,KawasakiCeIn3}. The case of
SrFe$_2$As$_2$ differs from any of examples known to date in that the SC and
AF orders are spatially distinguished but occur simultaneously forming a
spontaneous SC/AF hybrid structure.
\begin{acknowledgments}
We thank M. Ogata, Y. Yanase, K. Ishida, M. Yoshida, K. Matsubayashi and A.
 Yamada for their help and discussions. This work supported by the
 Grant-in-Aid for Scientific Research (B) (No. 21340093) from JSPS
 and by the GCOE program from MEXT.  K.K. is supported as a JSPS research fellow.
\end{acknowledgments}
\bibliography{document} 
\end{document}